\documentclass{aa} 
\usepackage{natbib}

\def\aap{{\it A\&A}}

\def\apjl{{\it ApJL}}

\begin{document}
\title{The complex gamma-ray behaviour of the Radio Galaxy M87}

\author{Faical Ait Benkhali
          \inst{1}
          \and
         Nachiketa Chakraborty\inst{1}
         \and
         Frank M. Rieger\inst{1,2}
         }
   \institute{$^1$ Max-Planck-Institut f\"ur Kernphysik, Saupfercheckweg 1, 69117 Heidelberg, Germany \\
              \email{F.Ait-Benkhali@mpi-hd.mpg.de} ; 
              \email{cnachi@mpi-hd.mpg.de}\\
             $^2$ ZAH, Institut f\"ur Theoretische Astrophysik, Universit\"at Heidelberg, Philosophenweg 12, 69120, Heidelberg, Germany \\
             \email{frank.rieger@mpi-hd.mpg.de}
             }

\abstract
{In recent years, non-blazar Active Galactic Nuclei (AGN) such as Radio Galaxies have emerged 
as a highly instructive source class providing unique insights into high energy acceleration 
and radiation mechanisms.} 
{Here we aim at a detailed characterization of the high-energy (HE; $>$100 MeV) gamma-ray 
emission from the prominent radio galaxy M87.} 
{We analyze $\sim8$ years of {\it Fermi}-LAT data and derive the spectral energy distribution 
between 100 MeV and 300 GeV. We extract lightcurves and investigate the variability behaviour 
for the entire energy range as well as below and above 10 GeV.}
{Our analysis provides (i) evidence for HE gamma-ray flux variability and (ii) indications for 
a possible excess over the standard power-law model above $E_b\sim10$ GeV, similar to the
earlier indications in the case of Cen A. 
When viewed in HE-VHE context, this is most naturally explained by an additional component 
dominating the highest-energy part of the spectrum. Investigation of the $\gamma$-ray lightcurves 
suggests that the lower-energy ($<10$ GeV) component is variable on timescales of (at least) a 
few months. The statistics of the high energy component ($>10$ GeV) does not allow significant 
constraints on variability. We find indications, however, for spectral changes with time that 
support variability of the putative additional component and seem to favor jet-related scenarios 
for its origin capable of accommodating month-type variability.} 
{The current findings suggest that both the high-energy ($> E_b$) and the very high energy (VHE; 
$>100$ GeV) emission in M87 are compatible with originating from the same physical component. 
The variability behaviour at VHE then allows further constraints on the location and the nature 
of the second component. In particular, these considerations suggest that the VHE emission during 
the quiescent state originates in a similar region as during the flare.}

\maketitle

\section{Introduction}\label{sect:intro}
The Giant Elliptical Virgo Cluster Galaxy M87 (NGC 4486, 3C274) is known for its prominent 
one-sided jet extending up to kiloparsec scales and for its extraordinary nuclear activity 
manifesting itself across all wavebands up to TeV energies \citep[e.g.,][]{2002ApJ...564..683M,
2007ApJ...668L..27K,2012MPLA...2730030R}. Being the second nearest \citep[distance $d\simeq 16.4
\pm0.5$,][]{2010A&A...524A..71B} active galaxy, M87 has been commonly classified as a 
low-excitation, weak-power Fanaroff-Riley (FR) I radio galaxy with a jet considered to be 
misaligned by $i\sim (15^o-25^o)$ \citep[e.g.,][]{2009Sci...325..444A}. Despite of hosting a very massive 
black hole (BH) in the range of $M_{BH} \simeq (2-6)\times 10^9 M_{\odot}$ \citep{1997MNRAS.289L..21M,2009ApJ...700.1690G,2013ApJ...770...86W}, 
M87 has been found to be highly underluminous suggesting that accretion onto its BH proceeds in a 
radiatively inefficient mode \citep[e.g.,][]{1996MNRAS.283L.111R}. Given its proximity 
and large gravitational reference scale, $r_g=GM_{BH}/c^2=6\times 10^{14} \;(4\times 
10^9 M_{\odot}/M_{BH})$ cm, M87 has been the focus of numerous observational campaigns enabling 
to study astrophysical processes such as e.g. jet formation or the production of non-thermal radiation 
in exceptional detail \citep[e.g.,][]{2012Sci...338..355D,2012MPLA...2730030R,2014ApJ...783L..33K,
2014ApJ...788..165H,2016A&A...595A..54M,2016ApJ...817..131H}.\\
M87 was the first extragalactic source detected at very high energies (VHE; $>100$ GeV). It is well
known for showing rapid (day-scale) VHE variability during active states and a hard, featureless spectrum 
(power-law photon index $\Gamma = 2.2 \pm 0.2 $ in high, and somewhat steeper $\Gamma\sim 2.6$ in 
low states) extending up $\sim 10$ TeV.  \citep{2003A&A...403L...1A,2006Sci...314.1424A,2008ApJ...685L..23A,
2009Sci...325..444A,2012ApJ...746..141A,2012ApJ...746..151A}. 
At high energies (HE; $>100$ MeV), {\it Fermi}-LAT has reported the detection of gamma-ray emission 
from M87 up to 30 GeV based on the first 10 months of LAT data \citep{2009ApJ...707...55A}, with a 
photon spectrum then compatible with a single power-law of index $\Gamma=2.26 \pm 0.13$, i.e. similar to the 
one(s) in the VHE high states. Despite these similarities, a simple extrapolation of this HE power-law 
to the VHE regime turns out to be insufficient to account for the flux levels measured during the TeV high states \citep{2012MPLA...2730030R}.
The HE lightcurve (on 10\,d bins) for these early observations does not show evidence for significant flux 
variations, though the occurrence of shorter-timescale variations cannot be excluded.  
The updated third {\it Fermi}-LAT Point Source Catalog (3FGL) based on 4 years of Pass 7 reprocessed 
data gives similar results with the spectrum below 10 GeV compatible with a single power-law of index $\Gamma
=2.04\pm0.07$ \citep{2015ApJS..218...23A}, yet suggesting a possible turnover above 10 GeV. 
The degree to which the TeV flux levels can be matched is thus an open issue. Our current work seeks to 
fill this gap using extended data sets. It is partly motivated by recent findings of an uncommon HE spectral 
hardening in the related source Cen~A  \citep{2013ApJ...770L...6S,2017PhRvD..95f3018B}.\\
A model comparison of the spectral energy distribution (SED) in M87 indicates that a one-zone synchrotron 
self-Compton (SSC) approach, in which the non-thermal emission processes in radio galaxies are taken to 
be similar to that in BL Lacs but with modest Doppler boosting \citep{2001MNRAS.324L..33C} (misaligned BL
Lac type with Doppler factor $D=1/[\gamma_b (1-\beta\cos i)] < 4$ for M87, where $\gamma_b$ is the bulk flow 
Lorentz factor), is unable to account for its overall SED up to TeV energies (including the high, and evidently 
also, the low states), yet capable of reproducing the emission below $\sim 10$ GeV \citep[e.g.,][]
{2009ApJ...707...55A}. This suggests that part of the HE-VHE emission may originate from a different (additional)
region. The analysis presented here casts new light on this.\\
Sections~2 and 3 describe the results of a general HE spectrum and lightcurve analysis of $\sim$8 year of 
{\it Fermi}-LAT data, while Sect.~4 explores the evidence for different spectral states. The results are discussed in Sect~5.

\section{Observations and Analysis}

{\it Fermi}-LAT is a $e^{\pm}$ pair-conversion $\gamma$-ray detector sensitive to photons in the energy range 
from $\sim20$ MeV to more than $300$ GeV \citep{2009ApJ...697.1071A}. It primarily operates in 
survey mode and continuously scans the whole sky every three hours.\\
Our current analysis uses \textsc{Pass8} (\texttt{P8R2}) algorithms \citep{2013arXiv1303.3514A} and data 
covering about 7.7 years from August 4, 2008 to April 12, 2016. A binned maximum likelihood analysis 
\citep{1996ApJ...461..396M} with a bin size of $0.1^{\circ}$ was performed using the standard Fermi Science 
Tools\footnote[1]{http://fermi.gsfc.nasa.gov/ssc/data/analysis/software} v10r0p5 software package.  
Events were selected with energies between 100 MeV and 300 GeV and in a region of interest (ROI) of $20^{\circ}$ 
around the position of M87 (RA=187.705, Dec=12.3911) with P8R2\_SOURCE\_V6 instrument response functions 
(IRF). We performed standard quality cuts in accordance with \textsc{Pass8} data analysis criteria. 
The background emission was modeled using the Galactic and isotropic diffuse emission \texttt{gll\_iem\_v06} 
and \texttt{iso\_P8R2\_SOURCE\_V6\_v06} files\footnote[2]{http://fermi.gsfc.nasa.gov/ssc/data/access/lat/BackgroundModels.html}.
All sources from the third LAT source catalog \cite[3FGL;][]{2015ApJS..218...23A} within the ROI are included 
in the model to ensure a satisfactory background modeling. Specifically, for the bright sources 3FGL J1224.9+2122, 
3FGL J1229.1+0202 and 3FGL J1239.5+0443 in the ROI, the spectral indices are allowed to vary in addition to 
varying the normalisations as for the other sources. Additionally photons coming from zenith angles larger than $90^{\circ}$ 
were all rejected to reduce the background from $\gamma$-rays produced in the atmosphere of the Earth (albedo). The 
Test Statistic value (TS), defined as $TS= -2\ln(L_0/L_1)$ was then used to determine the source detection 
significance, with threshold set to $TS=25$ ($\sim5\sigma$).\\ 
Our likelihood analysis reveals a point source with a high statistical significance $TS \simeq 1088$ ($\sim 33\sigma$).\\ 
The appearance of new sources within the ROI could in principle strongly influence the $\gamma$-ray spectrum and lightcurve 
of the investigated source. In order to investigate this we also searched for possible new $\gamma$-ray sources within a 
FOV of $20^{\circ}$. Our analysis of 8 year of Fermi-LAT data does not show any evidence of additional transient sources 
beyond the 3FGL catalog.\\ 
We produced the $\gamma$-ray SED and light curves of M87 through the unbinned and binned maximum likelihood fitting technique, 
respectively with \textit{gtlike} to determine the flux and TS value for each time bin. Both energy and temporal bins 
with $TS < 9$ (or $\sim 3\sigma$) are set as upper limits throughout the paper.
The effect of energy dispersion below 300 MeV is accounted for in the analysis by enabling the energy dispersion 
correction.

\section{The high energy gamma-ray SED of M87}
\label{sect:sedcomp}
The HE emission of M87 reveals hints of a deviation from a simple power law in the 
first 10 months of LAT data \citep{2009ApJ...707...55A} while the four-year (3FGL) 
update indicates a possible downturn at energies $\sim10$ GeV \citep{2015ApJS..218...23A}. 
Our current analysis based on $\approx$ 8 years of LAT data provides additional indications 
for a deviation from a single power law and excess emission above $\gtrsim$ 10 GeV.

\begin{figure}[htbp]
\begin{center}
\includegraphics[width=0.5\textwidth]{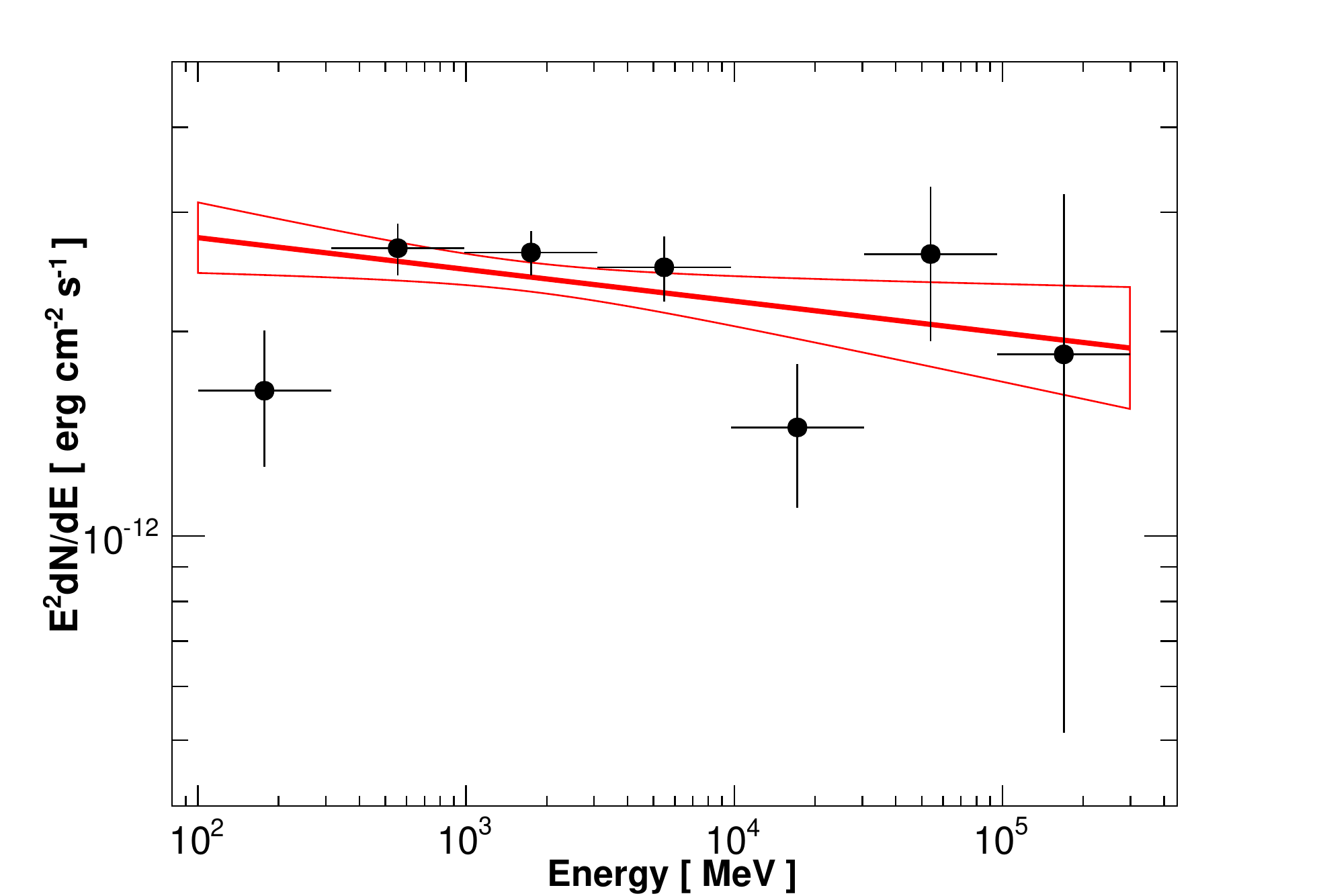}   
\caption{The SED of M87 in the energy band of 100 MeV to 300 GeV as extracted from the full (2008-2016) data set 
along with a power law fit. There are indications for an inflexion point around $\sim(5-30)$ GeV with additional 
spectral hardening towards higher energies, that appears suggestive of an additional emission component.}
\label{fig:fullSED}
\end{center}
\end{figure}
The SED for the full time and energy range (using 7 energy bins), is shown in Fig.~\ref{fig:fullSED}. 
 
\begin{figure}[htbp]
\begin{center}
\includegraphics[width=0.5\textwidth]{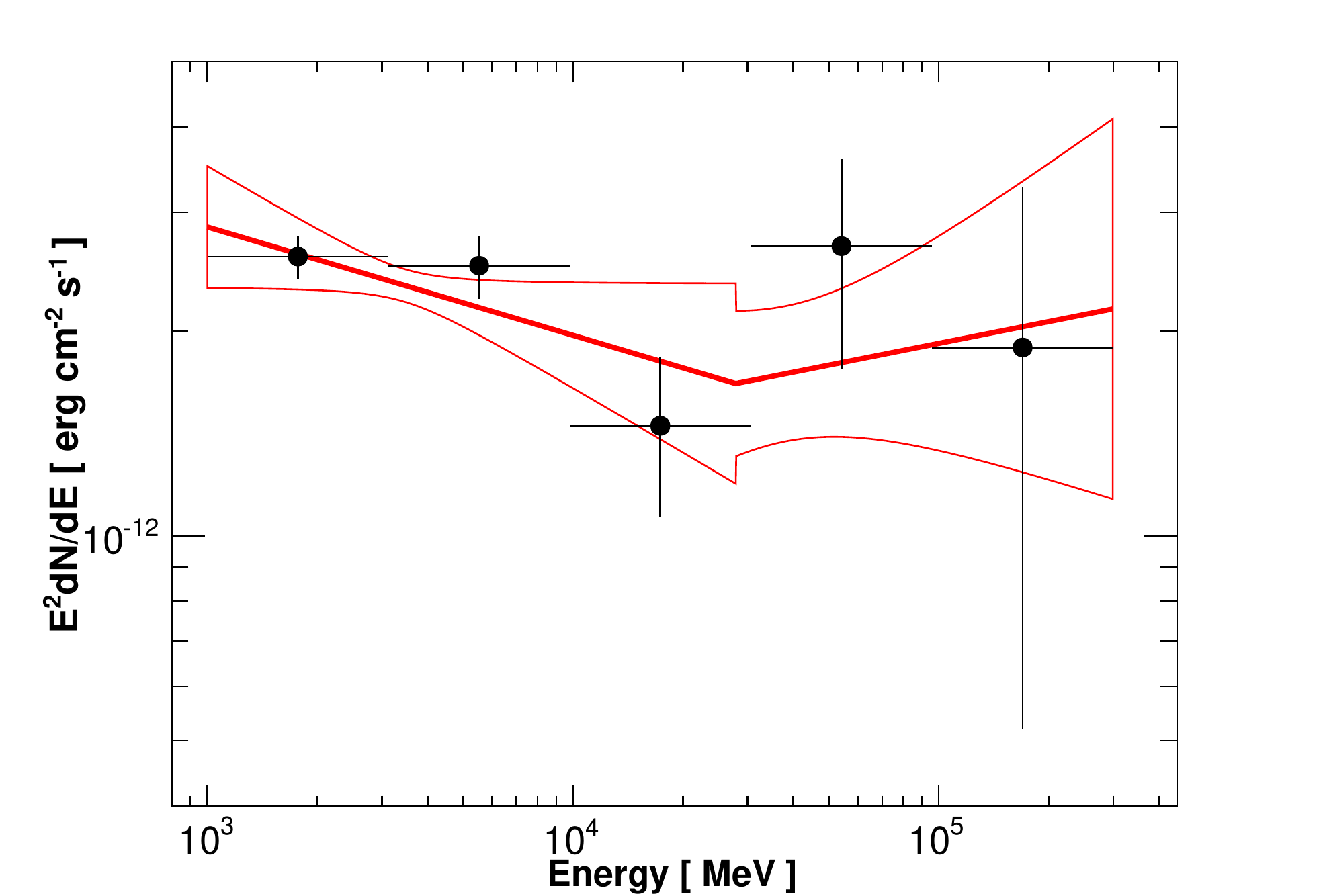}    
\caption{\bf The SED of M87 above 1 GeV for the full (2008-2016) data set along with the preferred 
broken power law fit.}
\label{fig:SED_high}
\end{center}
\end{figure}

We performed a maximum likelihood analysis exploring different spectral forms for the whole energy range, namely single 
power-law (PL),  
broken power-law (BPL) and log-parabola (LP). The likelihood ratio test comparison yields $TS=-2 \log(L_{PL}/L_{BPL}) \simeq 7.1$, thus preferring BPL 
over PL at a significance level of $\sim 2.66\ \sigma$. The best fit indices are $\Gamma_1=1.79\pm 0.15$ and 
$\Gamma_2 = 2.18\pm0.06$ with the break at $E_b = 1.32\pm0.34$ GeV. Here, $\Gamma_2 > \Gamma_1$ due to the 
curvature below 1 GeV, as evident from the $E^{2} dN / dE$ in Figure~\ref{fig:fullSED}. This is also hinted at by 
the preference of $2.1 \sigma$ of LP over PL; both are nested, so that the preference can be readily evaluated 
applying Wilks' theorem.\\
To minimize the influence of the curvature seen at lower and to study the spectral extension towards higher energies, 
we also fit the SED points above $1$ GeV. The results give a preference for a BPL over a PL at $2.4 \sigma$, with a break at 
$\sim 28\pm11$ GeV, and indices $\Gamma_1= 2.16\pm0.16$ and $\Gamma_2= 1.89 \pm0.29$, respectively (cf. 
Fig.~\ref{fig:SED_high}). A similar outome is obtained once $E_b$ is fixed separately to truely satisfy Wilks' theorem.
We note that this preference is comparable to the initial ($<3\sigma$) indications for spectral hardening at GeV energies 
in Cen~A \citep{2013ApJ...770L...6S}; given additional and more sensitive data, this preference has now increased to 
$\sim 5 \sigma$ \citep{2017PhRvD..95f3018B}. In the case of M87, the precise position of the possible break cannot be easily improved given available data as 
sufficient statistics necessitate wider bins. Nevertheless, the indications of a preference for a inverted BPL over a 
PL model above 1 GeV at $\gtrsim 2.5 \sigma$ suggests a non-trivial departure from the usual single component PL model. The outcome of an excess or inflexion above a few GeV critically depends upon the chosen bins at energies 5.5, 17.2 and 
54.0 GeV in Fig.~\ref{fig:fullSED}, each of which however is significant with TS values of 328, 70 and 56 respectively.\\ 

The above spectral analysis provides a hint for possible spectral inflexion 
at GeV energies. Such a situation might be expected if the source SED would 
actually be composed of more than one component. While not yet strong on its own, this hint 
gains plausibility when viewed in the wider HE-VHE context (see Sec.~\ref{sect:conclusions}). 
To experimentally improve the spectral characterization towards higher energies more sensitive 
data are needed, preferentially from lower threshold VHE observations by HESS, MAGIC and VERITAS 
or in time with CTA.

\section{Variability Analysis}
\label{sect:var}
\begin{figure}[htbp]
\begin{center}
\includegraphics[width=0.5\textwidth]{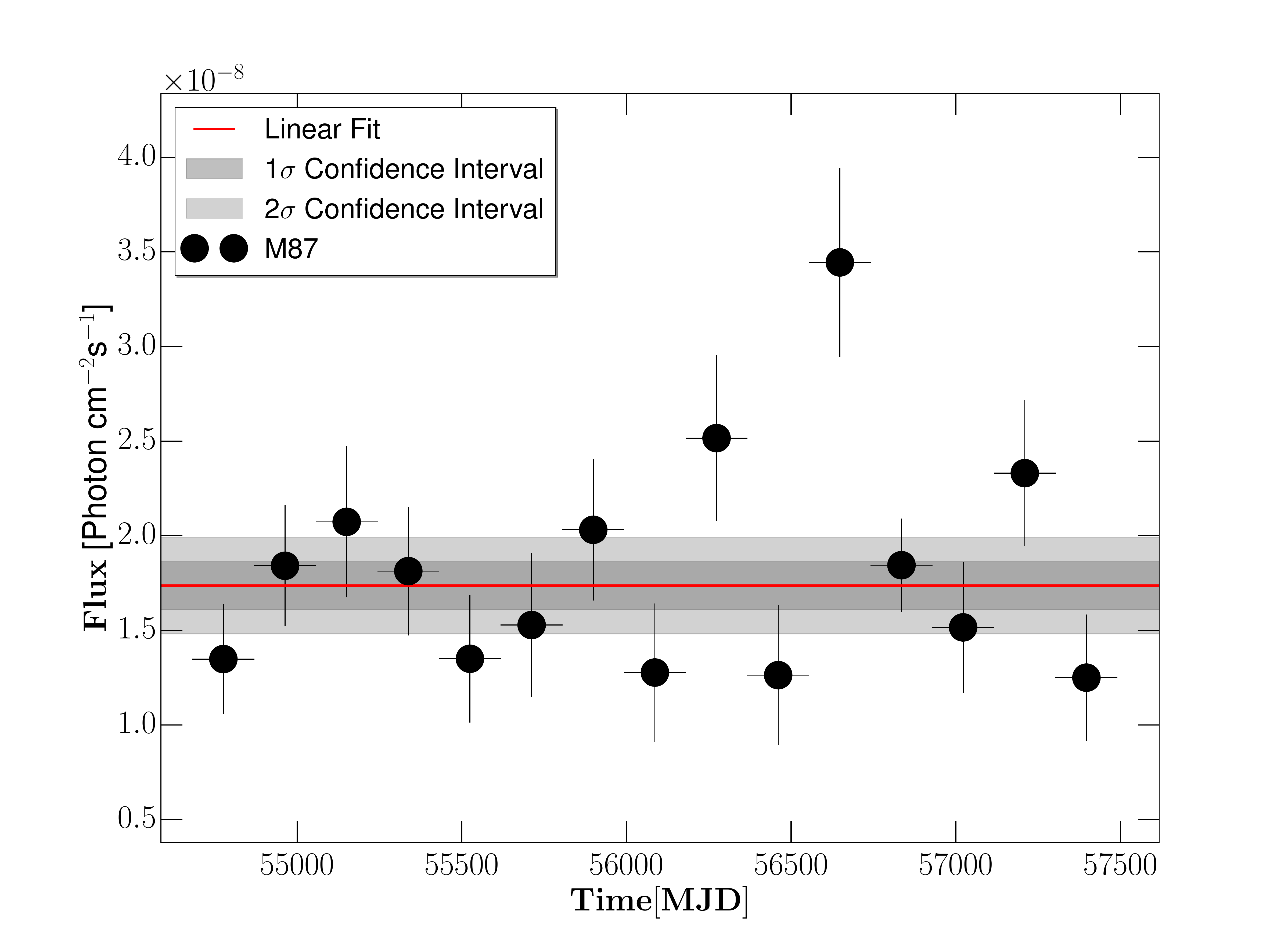}   
\caption{The $\gamma$-ray lightcurve of M87 from 2008 to 2016 in the energy range 0.1 - 300 GeV. 
A 6-month binning is employed. The red line gives a fit with a constant flux, while the 
shaded areas show the 1$\sigma$ and 2$\sigma$ confidence levels, respectively. There is clear evidence 
for variability. Error bars given are one-sigma error bars.}
\label{fig:LCfull}
\end{center}
\end{figure}

Early HE studies did not find evidence for significant variability during previous observations 
\citep{2009ApJ...707...55A,2012ApJ...746..141A}, though hints exist in the 3FGL lightcurve \citep{2015ApJS..218...23A}. 
On the other hand, at VHE energies M87 has shown at least three high flux states in 2005, 2008 and 2010 \citep[e.g.,][]{2006Sci...314.1424A,2008ApJ...685L..23A,2012ApJ...746..141A}, displaying extreme variability down to timescales of 
O($\sim$ 1 day). Given the extended data set with longer observation window and relatively better statistics, we 
investigate anew for possible HE variability. Figure~\ref{fig:LCfull} shows the full lightcurve with data from 2008 to 
2016 in the energy range 100 MeV to 300 GeV using a binning of six months. Clear variations of a factor of $\sim2$ over 
timescales of few months are apparent in the lightcurve. Fitting a constant yields $\chi^{2} / \nu = 28.1 / 14 $ with 
a probability $p< 0.014$ for being constant ($\sim 2.5 \sigma$). The 6-month binning ensures an optimal choice 
giving both sufficiently high statistics as well as a sufficiently high number of HE data points to investigate  
variations on timescales of several months. Given the weakness of the source, shorter-term variations are in 
general difficult to probe.

\begin{figure}[htbp]
\begin{center}
\includegraphics[width=0.49\textwidth]{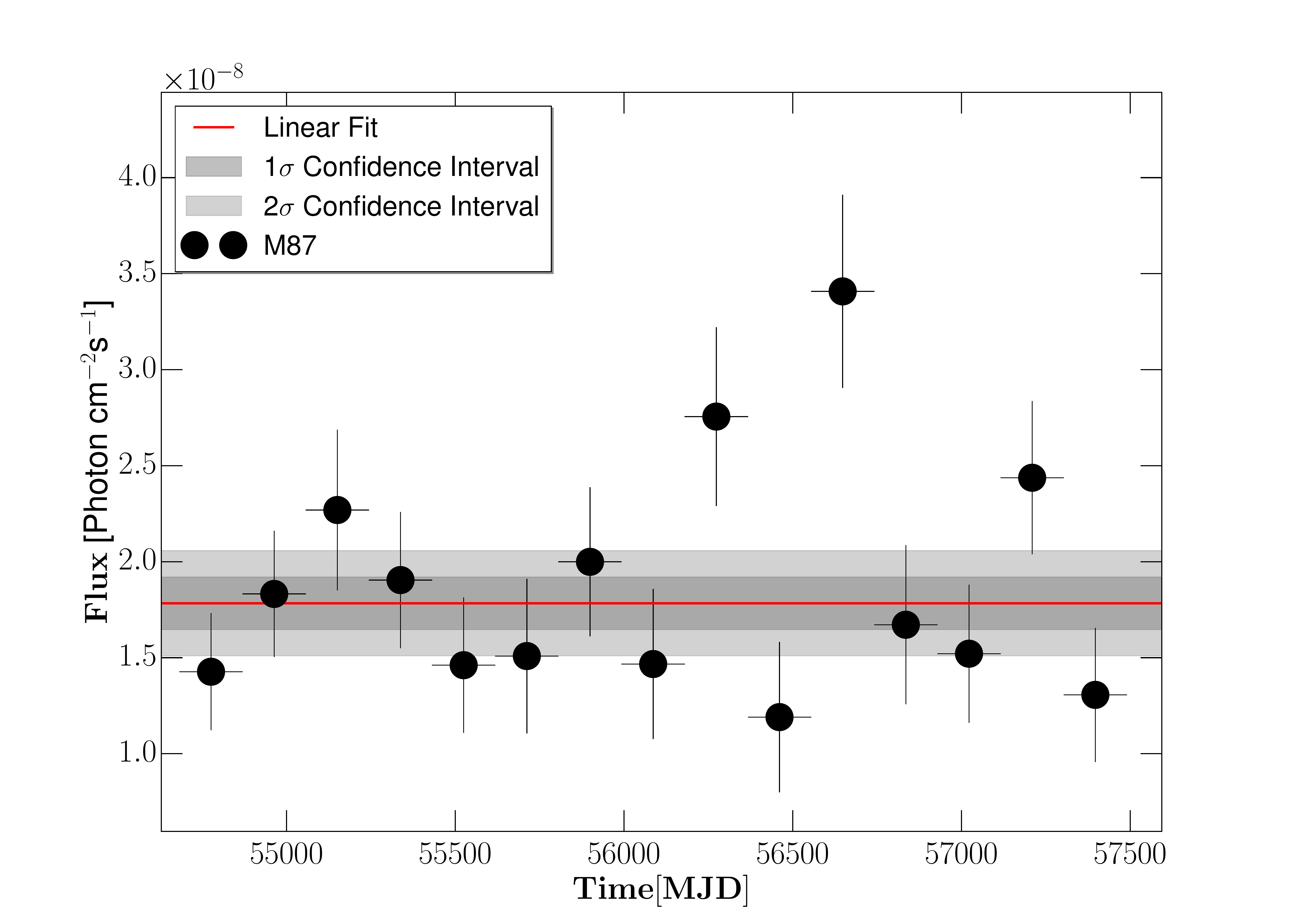} 
\includegraphics[width=0.49\textwidth]{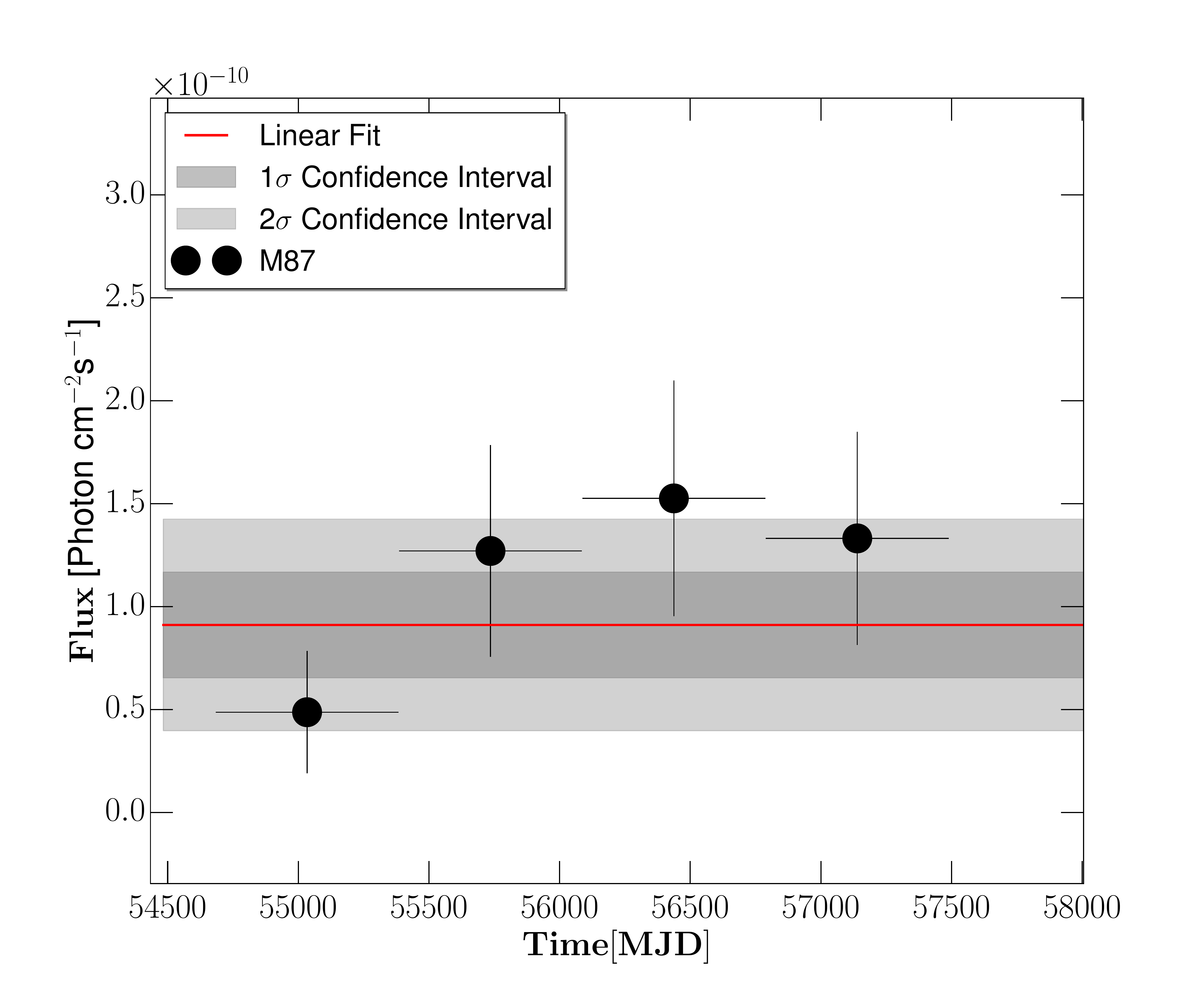} 
\caption{{\it Top}: The $\gamma$-ray lightcurve of M87 from 2008 to 2016 in the energy range 100 MeV - 10 GeV. The binning here is 6 months. There are clear deviations from a constant flux level. {\it Bottom}: The lightcurve in 
the energy band 10 GeV - 300 GeV with a $\sim2$ yr-binning. Here the statistics are not sufficient to establish any 
variability. Error bars given are one-sigma error bars.}
\label{fig:LCenergyband}
\end{center}
\end{figure}

The considerations in Sect.~\ref{sect:sedcomp} suggest a possible spectral inflexion somewhere 
between $5$ and $30$ GeV. Motivated by the spectral shape in Fig.~\ref{fig:fullSED}, the lightcurve is split into 
two energy bands, (0.1 - 10) GeV and (10 - 300) GeV, to investigate variability above and below a potential break 
or inflexion point $E_b$:\\ 
(i) For the first band, between 0.1-10 GeV, we again employ a 6-month binning, leading to 15 bins as for the 
full band. The corresponding lightcurve shows variability on similar timescales as before (see Fig.~\ref{fig:LCenergyband}). 
This is expected as the dominant statistical contribution to the full lightcurve comes from this part. Fitting a constant 
to this band results in $\chi^{2} =27.32$ for 14 degrees of freedom with p-value $\sim 0.018$. This confirms for the 
first time that the HE emission below 10 GeV is variable at least on timescales of a few months.\\ 
(ii) Above 10 GeV, on the other hand, we find that the 6-month binning yields several upper limits due to limited 
statistics and hence we use a 24 month binning ensuring that there are no upper limits. A constant fit yields $\chi^{2} 
= 4.33$ for 3 degrees of freedom with p-value 
$\sim 0.23$. Thus, variability in the high energy band $\gtrsim 10 $ GeV is not statistically significant and would 
require higher statistics per bin and larger number of bins to be clearly ruled out or established. Similar to the SED, 
this deserves further investigation with lower threshold TeV observations.\\ 
The evidence for variability below 10 GeV seems consistent with e.g. expectations based on one-zone SSC modeling of its 
SED up to the GeV regime. Detection of differences in the variability behaviour above and below $E_b$ could in principle 
provide further evidence for multiple components.

\section{Changes in Spectral State}
The spectral state shown in Fig.~\ref{fig:fullSED} is a long-term average SED of all the data from 2008 to 2016. This interval is known to encompass different source states at other energies, such as the major TeV flare recorded in 2010 
\citep{2012ApJ...746..151A,2012ApJ...746..141A}. The HE lightcurve analysis reported in Sect.~\ref{sect:var} 
provided clear signs for a slow (monthly) variability below $E_b \sim 10$ GeV, yet nothing conclusive for the regime 
above $E_b$. It seems interesting to explore whether an analysis of the spectrum over time allows further insights 
into the possible variability behaviour of the two components. Inspection of the lightcurve below $E_b$ reveals a 
rather high flux state in a single bin extending from September 2013 to March 2014. 
To study its impact, we split the whole dataset into this "high state" subset consisting of 6 months of data, and 
a subset called "regular state" consisting of the remaining data from 2008-2016. Note that the "regular state" in HE 
includes a strong short VHE flaring episode during April 5-15, 2010 \citep{2012ApJ...746..141A}. 
Further subdivision of the current dataset into intermediate states analogous to the TeV states remains 
non-conclusive given that the HE statistics is limited and that the GeV and TeV observations of the high states 
are not simultaneous. A Bayesian Blocks representation, that partitions the lightcurve into piecewise 
constant blocks by optimizing a fitness function \citep{2013ApJ...764..167S}, also does not favor further 
states when the standard geometric prior is employed.\\

\begin{figure}[htbp]
\begin{center}
\includegraphics[width=0.52\textwidth]{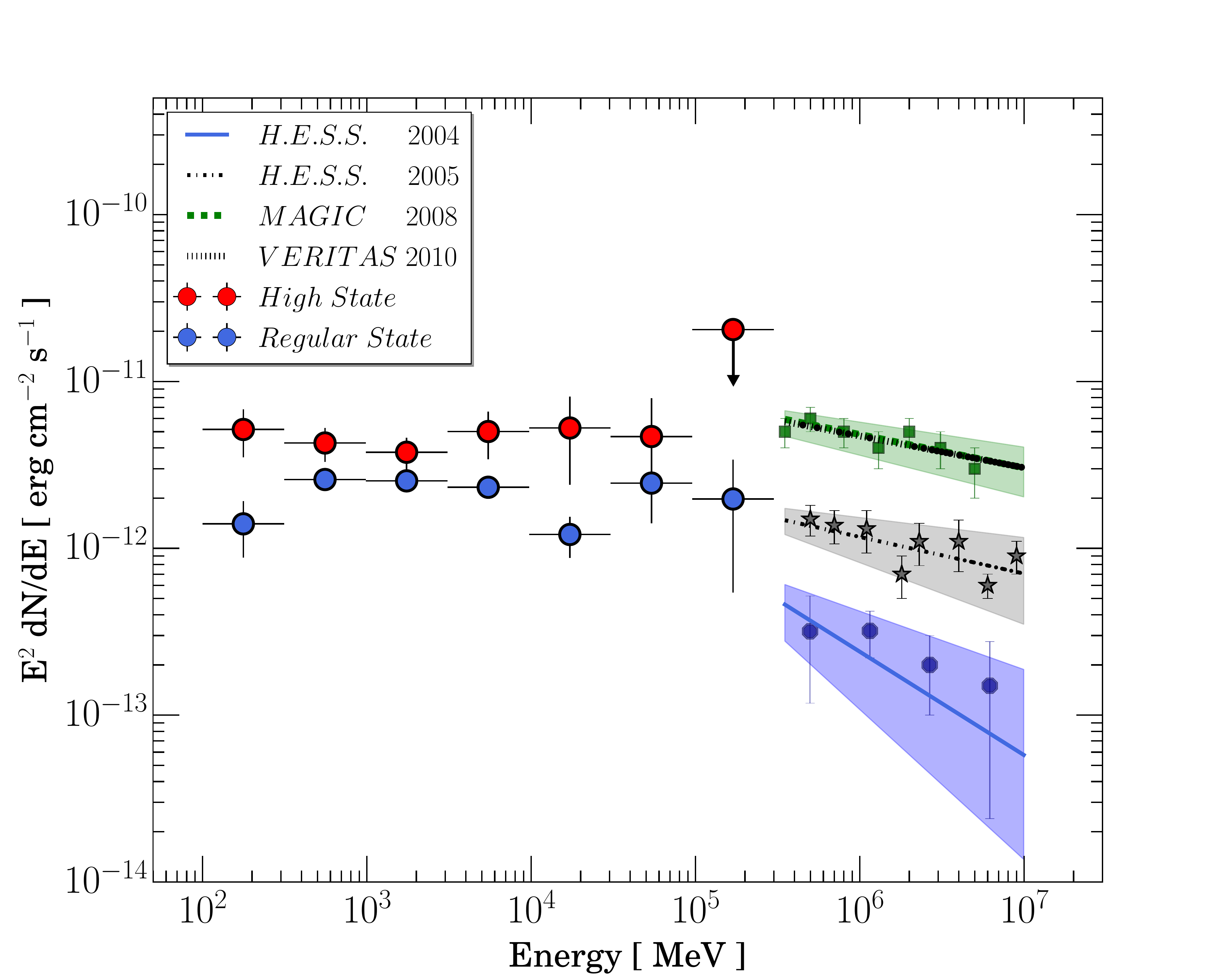}
\caption{The $\gamma$-ray SED of M87 for the regular and the high HE source state, respectively. 
The "high state" encompasses {\it Fermi}-LAT data from 09/2013 - 03/2014 and the "regular state" refers to 
the remaining data from 2008-2016. This "regular state" shows a break in the SED around $E\sim(5-30)$ 
GeV, as in Fig.~\ref{fig:fullSED}, suggestive of an additional HE component. The break 
appears camouflaged in the "high state" by flaring above $\sim10$ GeV. 
Lines and shaded regions, including VHE data points, indicate published best fit 
power laws and confidence bands for previous VHE measurements (blue line: H.E.S.S. VHE 
low state in 2004 \citep{2006Sci...314.1424A}; black dash-dotted: H.E.S.S. VHE high state 
in 2005 \citep{2006Sci...314.1424A}; green dotted: MAGIC VHE high state/flaring episode 
in February 2008 \citep{2008ApJ...685L..23A}; black dotted: VERITAS VHE high state/flaring episode in April 2010 \citep{2012ApJ...746..141A}).
Despite non-simultaneity, the observed HE spectral variability would now in principle 
allow matching the HE-VHE states.}
\label{fig:SEDstates}
\end{center}
\end{figure}

The respective SEDs are shown in Figure~\ref{fig:SEDstates}. For the "regular state" an inflexion in the SED 
between $\sim(5-30)$ GeV seems again indicated. On comparing the broken power-law model with the simple power-law model, we find a preference for the former at the level of $2.55 \sigma$ for the regular state, whereas there's no significant preference for BPL in the high state.  
The "high state" SED instead shows no remarkable drop but a rather smooth continuation (the inflexion being 
camouflaged) compatible with a single power law. The above analysis suggests, however, that this continuation 
may be caused by a rather smooth superposition of two physically different components. In fact, the observed 
spectral changes and the HE-VHE connection support the notion that the second component, which dominates the 
regular-state emission above $E_b$ is variable and that its variability shapes the high-state SED. Though beyond 
the scope of the present paper, it seems interesting to explore possible spectral changes as a function of HE flux 
levels in more detail.

\section{Discussion and Conclusions}
\label{sect:conclusions}
The analysis of the whole dataset provides evidence for month-type variability and indications for a 
possible spectral turnover in the GeV regime $E_b\gtrsim 10$ GeV. The latter seems reminiscent of 
the spectral hardening seen in the nearest radio galaxy, Cen~A \citep{2013ApJ...770L...6S,2017PhRvD..95f3018B}.\\ 
The current findings are most naturally interpreted as an additional physical component beyond the conventional 
SSC jet contribution for a misaligned AGN. Our variability analysis indicates that the component dominating 
the low energies ($E \leq E_b $) varies by a factor $\sim 2$ on timescale $\Delta t$ of at least a few 
months, which seems compatible with a jet-related origin. While the analysis did not ascertain variability of 
the second component, the observed spectral changes suggest that this component is variable on comparable 
timescales. This would disfavour steady or extended $\gamma$-ray production scenarios for its origin, such as 
inverse Compton scattering in its kiloparsec-scale jet \citep{2011MNRAS.415..133H}, dark matter annihilation 
\citep[e.g.][]{2000PhRvD..61b3514B,2011EPJC...71.1815S,2015PhRvD..92d3510L} or cosmic-ray interactions in the 
interstellar medium \citep[e.g.][]{2003A&A...407L..73P}.\\  
The current results instead rather favour a second component on scales of the inner jet and below, i.e., 
on scales $d\lesssim 1300~(\Delta t/1~\mathrm{month})~r_g$ for a jet half width to lenght ratio of $0.1$. Its 
true physical nature however, cannot yet be unambiguously inferred and a variety of different interpretations 
appear possible. Proposals in the literature encompass scenarios wherein the gamma-ray emission in M87 is attributed 
to leptonic or hadronic processes, or a combination thereof. The former range from inverse Compton scattering 
in a decelerating flow profile \citep{2005ApJ...634L..33G} or a spine-shear topology \citep{2008MNRAS.385L..98T} 
via multiple SSC emission zones \citep{2008A&A...478..111L} or various reconnection-driven events \citep[e.g.][]
{2010MNRAS.402.1649G} up to dynamic processes in its rotating black hole magnetosphere \citep[e.g.][]{2007ApJ...671...85N,2008A&A...479L...5R,2011ApJ...730..123L}. 
Hadronic proposals, on the other hand, include $p\gamma$ and proton synchrotron processes \citep{2004A&A...419...89R} 
or jet-star induced pp-interactions \citep[e.g.][]{2012ApJ...755..170B}, as well as complex lepto-hadronic realizations 
thereof \citep[e.g.,][]{2011A&A...531A..30R,2016ApJ...830...81F}. Most of these scenarios introduce additional 
parameters and no longer aim at describing the overall low energy (sub-mm)-VHE core SED by a single zone as 
attempted in earlier SSC models. They can thus potentially be modified to accommodate the current findings.\\
The connection of the HE and VHE regimes may deserve further investigations in this regard as it may eventually 
allow to substantiate general challenges associated with each of these scenarios \citep[see][for a discussion]
{2012MPLA...2730030R}. 
In the VHE monitoring campaign by MAGIC during 2012 to 2015 (Dec-July, with $\sim40$h each year) no VHE flares were 
detected, although the VHE lightcurve above 300 GeV seems to provide some hints for (day-scale) flickering in 2013 
\citep{Bangale:2015fah}. Some caution is needed, though, as the visibility at the MAGIC site (from Dec-July) constrains 
comparison, whereas the {\it Fermi}-LAT sensitivity limits the detectability of short flares above $\sim10$ GeV (as 
might have been the case in April 2010). Nevertheless, if these findings hold up to closer scrutiny, it would suggest 
that (1) under normal circumstances, flux changes associated with the second component, do not necessarily imply strong 
changes at VHE $\gamma$-rays, i.e. during a common HE increase its $\gamma$-ray SED peak tends to get shifted to lower 
energies. (2) On the other hand, during extraordinary VHE high states a reverse trend occurs, i.e. the $\gamma$-ray peak of the second 
component gets shifted to higher energies. Clearly then, the cross-over regime between HE and VHE, i.e., between 
space-based {\it Fermi}-LAT and ground-based Cherenkov arrays is particularly interesting.\\ 
The current {\it Fermi}-LAT analysis allows extraction of (significant) spectral points up to $\sim100$ GeV. At VHE energies, MAGIC reported a 
threshold of $300$ GeV for the monitoring noted above, while a detection down to $150$ GeV is known for the 2008 high 
state \citep{2008ApJ...685L..23A}. With existing performance upgrades such as H.E.S.S. II \citep{2017A&A...600A..89H} 
full energy overlap should thus soon be achievable.\\ 
In a minimalist approach the fact that the present {\it Fermi}-LAT 
spectrum extends up to about 100 GeV without indications for a strong cut-off, and the case that current VHE thresholds 
reach down to about 200 GeV already leaves rather little room but to attribute both the HE emission above $E_b$ and the VHE 
emission to anything other than the same physical component. While the limited sensitivity of {\it Fermi}-LAT for a weak 
source such as M87 does not allow to probe deeper into the variability at high energies, ground-based Cherenkov instruments 
have found the VHE emission to be variable down to timescales of days \citep[e.g.,][]{2006Sci...314.1424A,2008ApJ...685L..23A, 2009Sci...325..444A,2012ApJ...746..141A}. Given the scenario of the same component in both energy regimes, the day-scale 
variability seen at VHE imposes a general constraint on the overall nature, size ($r\lesssim 5~[\Delta t_{VHE}/1~\mathrm{day}]~r_g$
for negligible Doppler boosting) and location of this component ($d \lesssim 100~r_g$). Unless further components are 
invoked, changes in the flux state of this component would then be responsible for the different states seen in the HE 
(above $E_b$) and VHE domain as noted above. These considerations would then also suggest that the VHE emission during 
the quiescent state originates in a similar region as during the flare, further favouring inner-jet type scenarios 
\citep[e.g.,][]{2009Sci...325..444A}. Hints for a possible longterm evolution of the "quiescent" VHE state 
\citep{2012AIPC.1505..586B} appear consistent with this.\\ 
The findings presented here provide some further illustration that nearby radio galaxies are carrying a prime potential to probe 
deeper into the physics and nature of AGN \citep[e.g.,][]{2017AIPC.1792}, motivating further VHE observations to truly disentangle
the origin of the emission seen.

\begin{center}
\textbf{Acknowledgements}
\end{center}
NC kindly acknowledges support by the AvH foundation and MPIK, and FMR by a DFG Heisenberg Fellowship (RI 1187/4-1). 
We thank Werner Hofmann and Luigi Tibaldo for insightful comments. We also thank the referee for the comments that were useful in improving the paper.

\bibliography{egb}

\begin{thebibliography}{48}
\expandafter\ifx\csname natexlab\endcsname\relax\def\natexlab#1{#1}\fi

\bibitem[{{Abdalla} {et~al.}(2017){Abdalla}, {Abramowski}, {Aharonian}, {Ait
  Benkhali}, {Akhperjanian}, {Andersson}, {Ang{\"u}ner}, {Arrieta}, {Aubert},
  \& et~al.}]{2017A&A...600A..89H}
{Abdalla}, H., {Abramowski}, A., {Aharonian}, F., {et~al.} 2017, \aap, 600, A89

\bibitem[{{Abdo} {et~al.}(2009){Abdo}, {Ackermann}, {Ajello}, {Atwood},
  {Axelsson}, {Baldini}, {Ballet}, {Barbiellini}, {Bastieri}, {Bechtol},
  {Bellazzini}, {Berenji}, {Blandford}, {Bloom}, {Bonamente}, {Borgland},
  {Bregeon}, {Brez}, {Brigida}, {Bruel}, {Burnett}, {Caliandro}, {Cameron},
  {Cannon}, {Caraveo}, {Casandjian}, {Cavazzuti}, {Cecchi}, {{\c C}elik},
  {Charles}, {Cheung}, {Chiang}, {Ciprini}, {Claus}, {Cohen-Tanugi},
  {Colafrancesco}, {Conrad}, {Costamante}, {Cutini}, {Davis}, {Dermer}, {de
  Angelis}, {de Palma}, {Digel}, {Donato}, {Silva}, {Drell}, {Dubois},
  {Dumora}, {Edmonds}, {Farnier}, {Favuzzi}, {Fegan}, {Finke}, {Focke},
  {Fortin}, {Frailis}, {Fukazawa}, {Funk}, {Fusco}, {Gargano}, {Gasparrini},
  {Gehrels}, {Georganopoulos}, {Germani}, {Giebels}, {Giglietto}, {Giommi},
  {Giordano}, {Giroletti}, {Glanzman}, {Godfrey}, {Grenier}, {Grondin},
  {Grove}, {Guillemot}, {Guiriec}, {Hanabata}, {Harding}, {Hayashida}, {Hays},
  {Horan}, {J{\'o}hannesson}, {Johnson}, {Johnson}, {Johnson}, {Johnson},
  {Kamae}, {Katagiri}, {Kataoka}, {Kawai}, {Kerr}, {Kn{\"o}dlseder}, {Kocian},
  {Kuss}, {Lande}, {Latronico}, {Lemoine-Goumard}, {Longo}, {Loparco}, {Lott},
  {Lovellette}, {Lubrano}, {Madejski}, {Makeev}, {Mazziotta}, {McConville},
  {McEnery}, {Meurer}, {Michelson}, {Mitthumsiri}, {Mizuno}, {Moiseev},
  {Monte}, {Monzani}, {Morselli}, {Moskalenko}, {Murgia}, {Nolan}, {Norris},
  {Nuss}, {Ohsugi}, {Omodei}, {Orlando}, {Ormes}, {Ozaki}, {Paneque},
  {Panetta}, {Parent}, {Pelassa}, {Pepe}, {Pesce-Rollins}, {Piron}, {Porter},
  {Rain{\`o}}, {Rando}, {Razzano}, {Reimer}, {Reimer}, {Reposeur}, {Ritz},
  {Rochester}, {Rodriguez}, {Romani}, {Roth}, {Ryde}, {Sadrozinski},
  {Sambruna}, {Sanchez}, {Sander}, {Saz Parkinson}, {Scargle}, {Sgr{\`o}},
  {Shaw}, {Smith}, {Smith}, {Spandre}, {Spinelli}, {Strickman}, {Suson},
  {Tajima}, {Takahashi}, {Tanaka}, {Taylor}, {Thayer}, {Thompson}, {Tibaldo},
  {Torres}, {Tosti}, {Tramacere}, {Uchiyama}, {Usher}, {Vasileiou}, {Vilchez},
  {Waite}, {Wang}, {Winer}, {Wood}, {Ylinen}, {Ziegler}, {Harris}, {Massaro},
  \& {Stawarz}}]{2009ApJ...707...55A}
{Abdo}, A.~A., {Ackermann}, M., {Ajello}, M., {et~al.} 2009, \apj, 707, 55

\bibitem[{{Abramowski} {et~al.}(2012){Abramowski}, {Acero}, {Aharonian},
  {Akhperjanian}, {Anton}, {Balzer}, {Barnacka}, {Barres de Almeida},
  {Becherini}, {Becker}, \& et~al.}]{2012ApJ...746..151A}
{Abramowski}, A., {Acero}, F., {Aharonian}, F., {et~al.} 2012, \apj, 746, 151

\bibitem[{{Acciari} {et~al.}(2009){Acciari}, {Aliu}, {Arlen}, {Bautista},
  {Beilicke}, {Benbow}, {Bradbury}, {Buckley}, {Bugaev}, {Butt}, \&
  et~al.}]{2009Sci...325..444A}
{Acciari}, V.~A., {Aliu}, E., {Arlen}, T., {et~al.} 2009, Science, 325, 444

\bibitem[{{Acero} {et~al.}(2015){Acero}, {Ackermann}, {Ajello}, {Albert},
  {Atwood}, {Axelsson}, {Baldini}, {Ballet}, {Barbiellini}, {Bastieri},
  {Belfiore}, {Bellazzini}, {Bissaldi}, {Blandford}, {Bloom}, {Bogart},
  {Bonino}, {Bottacini}, {Bregeon}, {Britto}, {Bruel}, {Buehler}, {Burnett},
  {Buson}, {Caliandro}, {Cameron}, {Caputo}, {Caragiulo}, {Caraveo},
  {Casandjian}, {Cavazzuti}, {Charles}, {Chaves}, {Chekhtman}, {Cheung},
  {Chiang}, {Chiaro}, {Ciprini}, {Claus}, {Cohen-Tanugi}, {Cominsky}, {Conrad},
  {Cutini}, {D'Ammando}, {de Angelis}, {DeKlotz}, {de Palma}, {Desiante},
  {Digel}, {Di Venere}, {Drell}, {Dubois}, {Dumora}, {Favuzzi}, {Fegan},
  {Ferrara}, {Finke}, {Franckowiak}, {Fukazawa}, {Funk}, {Fusco}, {Gargano},
  {Gasparrini}, {Giebels}, {Giglietto}, {Giommi}, {Giordano}, {Giroletti},
  {Glanzman}, {Godfrey}, {Grenier}, {Grondin}, {Grove}, {Guillemot}, {Guiriec},
  {Hadasch}, {Harding}, {Hays}, {Hewitt}, {Hill}, {Horan}, {Iafrate}, {Jogler},
  {J{\'o}hannesson}, {Johnson}, {Johnson}, {Johnson}, {Johnson}, {Kamae},
  {Kataoka}, {Katsuta}, {Kuss}, {La Mura}, {Landriu}, {Larsson}, {Latronico},
  {Lemoine-Goumard}, {Li}, {Li}, {Longo}, {Loparco}, {Lott}, {Lovellette},
  {Lubrano}, {Madejski}, {Massaro}, {Mayer}, {Mazziotta}, {McEnery},
  {Michelson}, {Mirabal}, {Mizuno}, {Moiseev}, {Mongelli}, {Monzani},
  {Morselli}, {Moskalenko}, {Murgia}, {Nuss}, {Ohno}, {Ohsugi}, {Omodei},
  {Orienti}, {Orlando}, {Ormes}, {Paneque}, {Panetta}, {Perkins},
  {Pesce-Rollins}, {Piron}, {Pivato}, {Porter}, {Racusin}, {Rando}, {Razzano},
  {Razzaque}, {Reimer}, {Reimer}, {Reposeur}, {Rochester}, {Romani},
  {Salvetti}, {S{\'a}nchez-Conde}, {Saz Parkinson}, {Schulz}, {Siskind},
  {Smith}, {Spada}, {Spandre}, {Spinelli}, {Stephens}, {Strong}, {Suson},
  {Takahashi}, {Takahashi}, {Tanaka}, {Thayer}, {Thayer}, {Thompson},
  {Tibaldo}, {Tibolla}, {Torres}, {Torresi}, {Tosti}, {Troja}, {Van Klaveren},
  {Vianello}, {Winer}, {Wood}, {Wood}, {Zimmer}, \& {Fermi-LAT
  Collaboration}}]{2015ApJS..218...23A}
{Acero}, F., {Ackermann}, M., {Ajello}, M., {et~al.} 2015, \apjs, 218, 23

\bibitem[{{Aharonian} {et~al.}(2003){Aharonian}, {Akhperjanian}, {Beilicke},
  {Bernl{\"o}hr}, {B{\"o}rst}, {Bojahr}, {Bolz}, {Coarasa}, {Contreras},
  {Cortina}, {Denninghoff}, {Fonseca}, {Girma}, {G{\"o}tting}, {Heinzelmann},
  {Hermann}, {Heusler}, {Hofmann}, {Horns}, {Jung}, {Kankanyan}, {Kestel},
  {Kohnle}, {Konopelko}, {Kornmeyer}, {Kranich}, {Lampeitl}, {Lopez}, {Lorenz},
  {Lucarelli}, {Mang}, {Meyer}, {Mirzoyan}, {Moralejo}, {Ona-Wilhelmi},
  {Panter}, {Plyasheshnikov}, {P{\"u}hlhofer}, {de los Reyes}, {Rhode},
  {Ripken}, {Rowell}, {Sahakian}, {Samorski}, {Schilling}, {Siems},
  {Sobzynska}, {Stamm}, {Tluczykont}, {Vitale}, {V{\"o}lk}, {Wiedner}, \&
  {Wittek}}]{2003A&A...403L...1A}
{Aharonian}, F., {Akhperjanian}, A., {Beilicke}, M., {et~al.} 2003, \aap, 403,
  L1

\bibitem[{{Aharonian} {et~al.}(2006){Aharonian}, {Akhperjanian}, {Bazer-Bachi},
  {Beilicke}, {Benbow}, {Berge}, {Bernl{\"o}hr}, {Boisson}, {Bolz}, {Borrel},
  {Braun}, {Brown}, {B{\"u}hler}, {B{\"u}sching}, {Carrigan}, {Chadwick},
  {Chounet}, {Coignet}, {Cornils}, {Costamante}, {Degrange}, {Dickinson},
  {Djannati-Ata{\"i}}, {Drury}, {Dubus}, {Egberts}, {Emmanoulopoulos},
  {Espigat}, {Feinstein}, {Ferrero}, {Fiasson}, {Fontaine}, {Funk}, {Funk},
  {F{\"u}{\ss}ling}, {Gallant}, {Giebels}, {Glicenstein}, {Goret},
  {Hadjichristidis}, {Hauser}, {Hauser}, {Heinzelmann}, {Henri}, {Hermann},
  {Hinton}, {Hoffmann}, {Hofmann}, {Holleran}, {Hoppe}, {Horns},
  {Jacholkowska}, {de Jager}, {Kendziorra}, {Kerschhaggl}, {Kh{\'e}lifi},
  {Komin}, {Konopelko}, {Kosack}, {Lamanna}, {Latham}, {Le Gallou},
  {Lemi{\`e}re}, {Lemoine-Goumard}, {Lenain}, {Lohse}, {Martin},
  {Martineau-Huynh}, {Marcowith}, {Masterson}, {Maurin}, {McComb}, {Moulin},
  {de Naurois}, {Nedbal}, {Nolan}, {Noutsos}, {Orford}, {Osborne}, {Ouchrif},
  {Panter}, {Pelletier}, {Pita}, {P{\"u}hlhofer}, {Punch}, {Ranchon},
  {Raubenheimer}, {Raue}, {Rayner}, {Reimer}, {Ripken}, {Rob}, {Rolland},
  {Rosier-Lees}, {Rowell}, {Sahakian}, {Santangelo}, {Saug{\'e}}, {Schlenker},
  {Schlickeiser}, {Schr{\"o}der}, {Schwanke}, {Schwarzburg}, {Schwemmer},
  {Shalchi}, {Sol}, {Spangler}, {Spanier}, {Steenkamp}, {Stegmann}, {Superina},
  {Tam}, {Tavernet}, {Terrier}, {Tluczykont}, {van Eldik}, {Vasileiadis},
  {Venter}, {Vialle}, {Vincent}, {V{\"o}lk}, {Wagner}, \&
  {Ward}}]{2006Sci...314.1424A}
{Aharonian}, F., {Akhperjanian}, A.~G., {Bazer-Bachi}, A.~R., {et~al.} 2006,
  Science, 314, 1424

\bibitem[{{Albert} {et~al.}(2008){Albert}, {Aliu}, {Anderhub}, {Antonelli},
  {Antoranz}, {Backes}, {Baixeras}, {Barrio}, {Bartko}, {Bastieri}, {Becker},
  {Bednarek}, {Berger}, {Bernardini}, {Bigongiari}, {Biland}, {Bock},
  {Bonnoli}, {Bordas}, {Bosch-Ramon}, {Bretz}, {Britvitch}, {Camara},
  {Carmona}, {Chilingarian}, {Commichau}, {Contreras}, {Cortina}, {Costado},
  {Covino}, {Curtef}, {Dazzi}, {De Angelis}, {De Cea del Pozo}, {de los Reyes},
  {De Lotto}, {De Maria}, {De Sabata}, {Delgado Mendez}, {Dominguez}, {Dorner},
  {Doro}, {Errando}, {Fagiolini}, {Ferenc}, {Fern{\'a}ndez}, {Firpo},
  {Fonseca}, {Font}, {Galante}, {Garc{\'{\i}}a L{\'o}pez}, {Garczarczyk},
  {Gaug}, {Goebel}, {Hayashida}, {Herrero}, {H{\"o}hne}, {Hose}, {Hsu},
  {Huber}, {Jogler}, {Kranich}, {La Barbera}, {Laille}, {Leonardo}, {Lindfors},
  {Lombardi}, {Longo}, {L{\'o}pez}, {Lorenz}, {Majumdar}, {Maneva},
  {Mankuzhiyil}, {Mannheim}, {Maraschi}, {Mariotti}, {Mart{\'{\i}}nez},
  {Mazin}, {Meucci}, {Meyer}, {Miranda}, {Mirzoyan}, {Mizobuchi}, {Moles},
  {Moralejo}, {Nieto}, {Nilsson}, {Ninkovic}, {Otte}, {Oya}, {Panniello},
  {Paoletti}, {Paredes}, {Pasanen}, {Pascoli}, {Pauss}, {Pegna},
  {Perez-Torres}, {Persic}, {Peruzzo}, {Piccioli}, {Prada}, {Prandini},
  {Puchades}, {Raymers}, {Rhode}, {Rib{\'o}}, {Rico}, {Rissi}, {Robert},
  {R{\"u}gamer}, {Saggion}, {Saito}, {Salvati}, {Sanchez-Conde}, {Sartori},
  {Satalecka}, {Scalzotto}, {Scapin}, {Schweizer}, {Shayduk}, {Shinozaki},
  {Shore}, {Sidro}, {Sierpowska-Bartosik}, {Sillanp{\"a}{\"a}}, {Sobczynska},
  {Spanier}, {Stamerra}, {Stark}, {Takalo}, {Tavecchio}, {Temnikov}, {Tescaro},
  {Teshima}, {Tluczykont}, {Torres}, {Turini}, {Vankov}, {Venturini}, {Vitale},
  {Wagner}, {Wittek}, {Zabalza}, {Zandanel}, {Zanin}, \&
  {Zapatero}}]{2008ApJ...685L..23A}
{Albert}, J., {Aliu}, E., {Anderhub}, H., {et~al.} 2008, \apjl, 685, L23

\bibitem[{{Aliu} {et~al.}(2012){Aliu}, {Arlen}, {Aune}, {Beilicke}, {Benbow},
  {Bouvier}, {Bradbury}, {Buckley}, {Bugaev}, {Byrum}, {Cannon}, {Cesarini},
  {Ciupik}, {Collins-Hughes}, {Connolly}, {Cui}, {Dickherber}, {Duke},
  {Errando}, {Falcone}, {Finley}, {Finnegan}, {Fortson}, {Furniss}, {Galante},
  {Gall}, {Godambe}, {Griffin}, {Grube}, {Guenette}, {Gyuk}, {Hanna}, {Holder},
  {Huan}, {Hughes}, {Hui}, {Humensky}, {Imran}, {Kaaret}, {Karlsson},
  {Kertzman}, {Kieda}, {Krawczynski}, {Krennrich}, {Lang}, {LeBohec},
  {Madhavan}, {Maier}, {Majumdar}, {McArthur}, {McCann}, {Moriarty},
  {Mukherjee}, {Nu{\~n}ez}, {Ong}, {Orr}, {Otte}, {Park}, {Perkins}, {Pichel},
  {Pohl}, {Prokoph}, {Quinn}, {Ragan}, {Reyes}, {Reynolds}, {Roache}, {Rose},
  {Ruppel}, {Saxon}, {Schroedter}, {Sembroski}, {{\c S}ent{\"u}rk}, {Skole},
  {Staszak}, {Te{\v s}i{\'c}}, {Theiling}, {Thibadeau}, {Tsurusaki}, {Tyler},
  {Varlotta}, {Vassiliev}, {Vincent}, {Vivier}, {Wakely}, {Ward}, {Weekes},
  {Weinstein}, {Weisgarber}, {Williams}, \& {Zitzer}}]{2012ApJ...746..141A}
{Aliu}, E., {Arlen}, T., {Aune}, T., {et~al.} 2012, \apj, 746, 141

\bibitem[{{Atwood} {et~al.}(2013){Atwood}, {Albert}, {Baldini}, {Tinivella},
  {Bregeon}, {Pesce-Rollins}, {Sgr{\`o}}, {Bruel}, {Charles}, {Drlica-Wagner},
  {Franckowiak}, {Jogler}, {Rochester}, {Usher}, {Wood}, {Cohen-Tanugi}, \&
  {S.~Zimmer for the Fermi-LAT Collaboration}}]{2013arXiv1303.3514A}
{Atwood}, W., {Albert}, A., {Baldini}, L., {et~al.} 2013, ArXiv e-prints

\bibitem[{{Atwood} {et~al.}(2009){Atwood}, {Abdo}, {Ackermann}, {Althouse},
  {Anderson}, {Axelsson}, {Baldini}, {Ballet}, {Band}, {Barbiellini}, \&
  et~al.}]{2009ApJ...697.1071A}
{Atwood}, W.~B., {Abdo}, A.~A., {Ackermann}, M., {et~al.} 2009, \apj, 697, 1071

\bibitem[{{Baltz} {et~al.}(2000){Baltz}, {Briot}, {Salati}, {Taillet}, \&
  {Silk}}]{2000PhRvD..61b3514B}
{Baltz}, E.~A., {Briot}, C., {Salati}, P., {Taillet}, R., \& {Silk}, J. 2000,
  \prd, 61, 023514

\bibitem[{Bangale {et~al.}(2016)Bangale, Manganaro, Schultz, Colin, \&
  Mazin}]{Bangale:2015fah}
Bangale, P., Manganaro, M., Schultz, C., Colin, P., \& Mazin, D. 2016, PoS,
  ICRC2015, 759

\bibitem[{{Barkov} {et~al.}(2012){Barkov}, {Bosch-Ramon}, \&
  {Aharonian}}]{2012ApJ...755..170B}
{Barkov}, M.~V., {Bosch-Ramon}, V., \& {Aharonian}, F.~A. 2012, \apj, 755, 170

\bibitem[{{Beilicke} \& {et al.}(2012)}]{2012AIPC.1505..586B}
{Beilicke}, M., \& {et al.} 2012, in AIP Conference Series, Vol. 1505, High
  Energy Gamma-Ray Astronomy, ed. F.~A. {Aharonian}, W.~{Hofmann}, \& F.~M.
  {Rieger}, 586--589

\bibitem[{{Bird} {et~al.}(2010){Bird}, {Harris}, {Blakeslee}, \&
  {Flynn}}]{2010A&A...524A..71B}
{Bird}, S., {Harris}, W.~E., {Blakeslee}, J.~P., \& {Flynn}, C. 2010, \aap,
  524, A71

\bibitem[{{Brown} {et~al.}(2017){Brown}, {B\"ohm}, {Graham}, {Lacroix},
  {Chadwick}, \& {Silk}}]{2017PhRvD..95f3018B}
{Brown}, A.~M., {B\"ohm}, C., {Graham}, J., {et~al.} 2017, \prd, 95, 063018

\bibitem[{{Chiaberge} {et~al.}(2001){Chiaberge}, {Capetti}, \&
  {Celotti}}]{2001MNRAS.324L..33C}
{Chiaberge}, M., {Capetti}, A., \& {Celotti}, A. 2001, \mnras, 324, L33

\bibitem[{{Doeleman} {et~al.}(2012){Doeleman}, {Fish}, {Schenck}, {Beaudoin},
  {Blundell}, {Bower}, {Broderick}, {Chamberlin}, {Freund}, {Friberg},
  {Gurwell}, {Ho}, {Honma}, {Inoue}, {Krichbaum}, {Lamb}, {Loeb}, {Lonsdale},
  {Marrone}, {Moran}, {Oyama}, {Plambeck}, {Primiani}, {Rogers}, {Smythe},
  {SooHoo}, {Strittmatter}, {Tilanus}, {Titus}, {Weintroub}, {Wright}, {Young},
  \& {Ziurys}}]{2012Sci...338..355D}
{Doeleman}, S.~S., {Fish}, V.~L., {Schenck}, D.~E., {et~al.} 2012, Science,
  338, 355

\bibitem[{{Fraija} \& {Marinelli}(2016)}]{2016ApJ...830...81F}
{Fraija}, N., \& {Marinelli}, A. 2016, \apj, 830, 81

\bibitem[{{Gebhardt} \& {Thomas}(2009)}]{2009ApJ...700.1690G}
{Gebhardt}, K., \& {Thomas}, J. 2009, \apj, 700, 1690

\bibitem[{{Georganopoulos} {et~al.}(2005){Georganopoulos}, {Perlman}, \&
  {Kazanas}}]{2005ApJ...634L..33G}
{Georganopoulos}, M., {Perlman}, E.~S., \& {Kazanas}, D. 2005, \apjl, 634, L33

\bibitem[{{Giannios} {et~al.}(2010){Giannios}, {Uzdensky}, \&
  {Begelman}}]{2010MNRAS.402.1649G}
{Giannios}, D., {Uzdensky}, D.~A., \& {Begelman}, M.~C. 2010, \mnras, 402, 1649

\bibitem[{{Hada} {et~al.}(2014){Hada}, {Giroletti}, {Kino}, {Giovannini},
  {D'Ammando}, {Cheung}, {Beilicke}, {Nagai}, {Doi}, {Akiyama}, {Honma},
  {Niinuma}, {Casadio}, {Orienti}, {Krawczynski}, {G{\'o}mez}, {Sawada-Satoh},
  {Koyama}, {Cesarini}, {Nakahara}, \& {Gurwell}}]{2014ApJ...788..165H}
{Hada}, K., {Giroletti}, M., {Kino}, M., {et~al.} 2014, \apj, 788, 165

\bibitem[{{Hada} {et~al.}(2016){Hada}, {Kino}, {Doi}, {Nagai}, {Honma},
  {Akiyama}, {Tazaki}, {Lico}, {Giroletti}, {Giovannini}, {Orienti}, \&
  {Hagiwara}}]{2016ApJ...817..131H}
{Hada}, K., {Kino}, M., {Doi}, A., {et~al.} 2016, \apj, 817, 131

\bibitem[{{Hardcastle} \& {Croston}(2011)}]{2011MNRAS.415..133H}
{Hardcastle}, M.~J., \& {Croston}, J.~H. 2011, \mnras, 415, 133

\bibitem[{{Kovalev} {et~al.}(2007){Kovalev}, {Lister}, {Homan}, \&
  {Kellermann}}]{2007ApJ...668L..27K}
{Kovalev}, Y.~Y., {Lister}, M.~L., {Homan}, D.~C., \& {Kellermann}, K.~I. 2007,
  \apjl, 668, L27

\bibitem[{{Kuo} {et~al.}(2014){Kuo}, {Asada}, {Rao}, {Nakamura}, {Algaba},
  {Liu}, {Inoue}, {Koch}, {Ho}, {Matsushita}, {Pu}, {Akiyama}, {Nishioka}, \&
  {Pradel}}]{2014ApJ...783L..33K}
{Kuo}, C.~Y., {Asada}, K., {Rao}, R., {et~al.} 2014, \apjl, 783, L33

\bibitem[{{Lacroix} {et~al.}(2015){Lacroix}, {Boehm}, \&
  {Silk}}]{2015PhRvD..92d3510L}
{Lacroix}, T., {Boehm}, C., \& {Silk}, J. 2015, \prd, 92, 043510

\bibitem[{{Lenain} {et~al.}(2008){Lenain}, {Boisson}, {Sol}, \&
  {Katarzy{\'n}ski}}]{2008A&A...478..111L}
{Lenain}, J.-P., {Boisson}, C., {Sol}, H., \& {Katarzy{\'n}ski}, K. 2008, \aap,
  478, 111

\bibitem[{{Levinson} \& {Rieger}(2011)}]{2011ApJ...730..123L}
{Levinson}, A., \& {Rieger}, F. 2011, \apj, 730, 123

\bibitem[{{Marconi} {et~al.}(1997){Marconi}, {Axon}, {Macchetto}, {Capetti},
  {Sparks}, \& {Crane}}]{1997MNRAS.289L..21M}
{Marconi}, A., {Axon}, D.~J., {Macchetto}, F.~D., {et~al.} 1997, \mnras, 289,
  L21

\bibitem[{{Marshall} {et~al.}(2002){Marshall}, {Miller}, {Davis}, {Perlman},
  {Wise}, {Canizares}, \& {Harris}}]{2002ApJ...564..683M}
{Marshall}, H.~L., {Miller}, B.~P., {Davis}, D.~S., {et~al.} 2002, \apj, 564,
  683

\bibitem[{{Mattox} {et~al.}(1996){Mattox}, {Bertsch}, {Chiang}, {Dingus},
  {Digel}, {Esposito}, {Fierro}, {Hartman}, {Hunter}, {Kanbach}, {Kniffen},
  {Lin}, {Macomb}, {Mayer-Hasselwander}, {Michelson}, {von Montigny},
  {Mukherjee}, {Nolan}, {Ramanamurthy}, {Schneid}, {Sreekumar}, {Thompson}, \&
  {Willis}}]{1996ApJ...461..396M}
{Mattox}, J.~R., {Bertsch}, D.~L., {Chiang}, J., {et~al.} 1996, \apj, 461, 396

\bibitem[{{Mertens} {et~al.}(2016){Mertens}, {Lobanov}, {Walker}, \&
  {Hardee}}]{2016A&A...595A..54M}
{Mertens}, F., {Lobanov}, A.~P., {Walker}, R.~C., \& {Hardee}, P.~E. 2016,
  \aap, 595, A54

\bibitem[{{Neronov} \& {Aharonian}(2007)}]{2007ApJ...671...85N}
{Neronov}, A., \& {Aharonian}, F.~A. 2007, \apj, 671, 85

\bibitem[{{Pfrommer} \& {En{\ss}lin}(2003)}]{2003A&A...407L..73P}
{Pfrommer}, C., \& {En{\ss}lin}, T.~A. 2003, \aap, 407, L73

\bibitem[{{Reimer} {et~al.}(2004){Reimer}, {Protheroe}, \&
  {Donea}}]{2004A&A...419...89R}
{Reimer}, A., {Protheroe}, R.~J., \& {Donea}, A.-C. 2004, \aap, 419, 89

\bibitem[{{Reynolds} {et~al.}(1996){Reynolds}, {Di Matteo}, {Fabian}, {Hwang},
  \& {Canizares}}]{1996MNRAS.283L.111R}
{Reynolds}, C.~S., {Di Matteo}, T., {Fabian}, A.~C., {Hwang}, U., \&
  {Canizares}, C.~R. 1996, \mnras, 283, L111

\bibitem[{{Reynoso} {et~al.}(2011){Reynoso}, {Medina}, \&
  {Romero}}]{2011A&A...531A..30R}
{Reynoso}, M.~M., {Medina}, M.~C., \& {Romero}, G.~E. 2011, \aap, 531, A30

\bibitem[{{Rieger}(2017)}]{2017AIPC.1792}
{Rieger}, F.~M. 2017, in AIP Conference Series, Vol. 1792, High Energy
  Gamma-Ray Astronomy, ed. F.~A. {Aharonian}, W.~{Hofmann}, \& F.~M. {Rieger},
  020008

\bibitem[{{Rieger} \& {Aharonian}(2012)}]{2012MPLA...2730030R}
{Rieger}, F.~M., \& {Aharonian}, F. 2012, Modern Physics Letters A, 27, 1230030

\bibitem[{{Rieger} \& {Aharonian}(2008)}]{2008A&A...479L...5R}
{Rieger}, F.~M., \& {Aharonian}, F.~A. 2008, \aap, 479, L5

\bibitem[{{Sahakyan} {et~al.}(2013){Sahakyan}, {Yang}, {Aharonian}, \&
  {Rieger}}]{2013ApJ...770L...6S}
{Sahakyan}, N., {Yang}, R., {Aharonian}, F.~A., \& {Rieger}, F.~M. 2013, \apjl,
  770, L6

\bibitem[{{Saxena} {et~al.}(2011){Saxena}, {Summa}, {Els{\"a}sser},
  {R{\"u}ger}, \& {Mannheim}}]{2011EPJC...71.1815S}
{Saxena}, S., {Summa}, A., {Els{\"a}sser}, D., {R{\"u}ger}, M., \& {Mannheim},
  K. 2011, European Physical Journal C, 71, 1815

\bibitem[{{Scargle} {et~al.}(2013){Scargle}, {Norris}, {Jackson}, \&
  {Chiang}}]{2013ApJ...764..167S}
{Scargle}, J.~D., {Norris}, J.~P., {Jackson}, B., \& {Chiang}, J. 2013, \apj,
  764, 167

\bibitem[{{Tavecchio} \& {Ghisellini}(2008)}]{2008MNRAS.385L..98T}
{Tavecchio}, F., \& {Ghisellini}, G. 2008, \mnras, 385, L98

\bibitem[{{Walsh} {et~al.}(2013){Walsh}, {Barth}, {Ho}, \&
  {Sarzi}}]{2013ApJ...770...86W}
{Walsh}, J.~L., {Barth}, A.~J., {Ho}, L.~C., \& {Sarzi}, M. 2013, \apj, 770, 86

\end{thebibliography}
\end{document}